**Vain is the pursuit of gravity waves**

A. LOINGER

Dipartimento di Fisica, Università di Milano

Via Celoria, 16,  20133 Milano, Italy

SUMMARY. − The modern apparatuses for the detection of the gravity waves are devised with the purpose to exploit the geodesic deviation generated by them. But the *pseudo* energy-momentum of these waves cannot exert any physical action on the apparatuses.

PACS. 04.80 − Experimental tests of general relativity and observations of gravitational radiation;

04.30 − Gravitational waves and radiation: theory.

**1**. − In a recent interesting paper De Salvo has reviewed the state of the experimental quest for gravitational waves [1]. As is known, there are two kinds of apparatuses currently designed and employed: resonant bars and interferometric devices. The bars have an evident drawback: they can respond essentially to signals with frequencies very close to the bar ringing frequency (of the order of 1 kHz). However, a wide-band detection technique is supplied by Michelson interferometers with kilometric arms and mirrors suspended to threads "so as to be free to be jerked by incoming gravitational waves." [1]. The interferometers are composed of "Fabry-Perot cavities to trap the light for long periods and increase the sensitivity. Laser standing powers measured in kW will be stored within the cavities." [1]. After a detailed description of the subtle stratagems which are necessary to neutralize the various extraneous noises, De Salvo concludes with a note of moderate optimism. He writes: "Only a large reduction of readout noise will allow for real gravitational wave astronomy. This, along with the networking of many interferometers at large separations, will grant us at long last a peek into the optically dense regions of the universe (as, for example, the core of our own galaxy) where most of the mass resides. […] . The quest, like so many others, begins with an elusive goal fleetingly imagined. But with industry, intelligence and perseverance, it may produce a significant scientific trophy."

In a series of papers [2], I have pointed out that there are many reasons to think that the gravity waves are entities destitute of a real physical existence. I shall now discuss in a detailed way a direct argument which demonstrates that all the detection "mechanisms" of the gravitational waves are condemned to fail.

**2**. − Nowadays the great majority of the theoreticians has dismissed the naïve belief that any *accelerated* mass point must emit gravitational waves; also the faith in the actual value of the famous *quadrupole formula* has almost evaporated. And for good reasons.

The research is now based on the *geodesic deviation*, the experimental devices are contrived to reveal the related effects. It is therefore advisable to re-examine the theoretical aspects of this phenomenon. Our analysis will be founded on the memoir "Sur l'écart géodésique" by Levi-Civita [3], the first investigator of the Riemannian geodesic deviation, which is unsurpassed in clarity, completeness and logical rigour.

**3**. − Levi-Civita [3] remarks, first of all, that we have a linear differential equation, due to Jacobi, that characterizes the set $I$ of the geodesics $\gamma$ of a (two-dimensional) surface $\Sigma$, which are infinitely close to a given geodesic $B$, named *base*. This classic Jacobi's formula is:

(3.1) $$\frac{d^2 y}{d\sigma^2} + K(\sigma)\, y = 0 \quad,$$

where: $y$ is the (infinitesimal) distance of a generic point $M$ of $\gamma$ from the base $B$; $\sigma$ is the curvilinear abscissa (with an arbitrary origin $O$) of the normal projection $P$ of $M$ on $B$; $K(\sigma)$ is the Gaussian curvature of $\Sigma$ at $P$.

In his paper Levi-Civita generalizes eq. (3.1) – and accordingly the concept of *geodesic deviation* $y$ – to a Riemannian manifold $V_n$, with any dimension $n$, characterized by a positive definite $ds^2$:

(3.2) $$ds^2 = a_{jk}\, dx^j\, dx^k \quad, (j, k = 1, 2, \ldots, n)\, .$$

However, the following results will be essentially valid also for an *indefinite* and *irreducible* form, under the assumption that *the length of the base B is different from zero*.

In a system of general co-ordinates $x^r$ ($r = 1, 2, \ldots, n$) the differential equations of a geodesic $\gamma$ of $V_n$ are, as is known:

(3.3) $$\ddot{x}^i = -\Gamma^i_{jh}\, \dot{x}^j\, \dot{x}^h \quad, \quad (\dot{x}^r \equiv dx^r/ds\, , \text{etc.})\, ,$$

if $\Gamma$ is the Christoffel symbol of second kind. We write

(3.4) $$x^i(s) = \varphi^i(\sigma) + \xi^i(\sigma)\ ;$$

the $\varphi^i(\sigma)$'s are the co-ordinates of the points of $B$; the $\xi^i(\sigma)$'s, and their derivatives with respect to $\sigma$, are first-order infinitesimals. Obviously $PM=(\xi^i)$. For generality's sake, we do *not* assume that the infinitesimal vector $PM$ is orthogonal to $B$: its orientation will depend on the *arbitrary* correspondence rule between the points $P$ and $M$ of the geodesics $\gamma$ and $B$. Moreover, $ds \neq d\sigma$; let us put

(3.5) $$\frac{ds}{d\sigma} = 1 + \lambda(\sigma) \ ;$$

it is not difficult to prove that

(3.6) $$\lambda(\sigma) = \frac{D\xi^r(\sigma)}{D\sigma} \dot{\varphi}_r(\sigma) \ ,$$

where: $\dot{\varphi}_r(\sigma) = a_{ir}\dot{\varphi}^i(\sigma) \equiv a_{ir}\dfrac{d\varphi^i}{d\sigma}$, and $D\xi^r/D\sigma$ is the intrinsic derivative defined by

(3.7) $$\frac{D\xi^r}{D\sigma} := \frac{d\xi^r}{d\sigma} + \Gamma^r{}_{ih}\dot{\varphi}^i\xi^h \ .$$

As it was intuitive, $\lambda(\sigma)$ is an *infinitesimal* scalar.

With some formal manipulations and bearing in mind the differential equations of the geodesics, we obtain

(3.8) $$\frac{d^2\xi^r}{d\sigma^2} - \frac{d\lambda}{d\sigma}\dot{\varphi}^r + 2\Gamma^r{}_{jh}\ \dot{\varphi}^j\frac{d\xi^h}{d\sigma} = -\Gamma^r{}_{ih,k}\dot{\varphi}^i\dot{\varphi}^h\xi^k \ ,$$

where the comma denotes the ordinary (not covariant) derivative. Then, by computing the intrinsic second derivative $D^2\xi^r/D\sigma^2$, we have

(3.9) $$\frac{D^2\xi^r}{D\sigma^2} = \frac{d^2\xi^r}{d\sigma^2} + 2\Gamma^r{}_{jh}\ \dot{\varphi}^j\frac{d\xi^h}{d\sigma} + \Xi^{(r)} \ ,$$

where

(3.10) $$\Xi^{(r)} \equiv \Gamma^r{}_{ik,h}\dot{\varphi}^i\dot{\varphi}^h\xi^k + \Gamma^r{}_{lk}\xi^k\ddot{\varphi}^l + \Gamma^r{}_{lh}\Gamma^l{}_{ik}\dot{\varphi}^i\dot{\varphi}^h\xi^k \ ;$$

now, the differential equations of $B$ tell us that

(3.11) $$\ddot{\varphi}^l = -\Gamma^l{}_{ih}\dot{\varphi}^i\dot{\varphi}^h \ ;$$

by substituting (3.11) in (3.10), adding $\Xi^{(r)}$ to both sides of (3.8), remembering the definition of the curvature tensor $R^r{}_{ihk}$

(3.12) $$R^r{}_{ihk} := \Gamma^r{}_{ih,k} - \Gamma^r{}_{ik,h} + \Gamma^r{}_{lk}\Gamma^l{}_{ih} - \Gamma^r{}_{lh}\Gamma^l{}_{ik} \; ,$$

eqs. (3.9) can be written as follows:

(3.13) $$\frac{D^2\xi^r}{D\sigma^2} = \frac{d\lambda}{d\sigma}\dot\varphi^r - R^r{}_{ihk}\,\dot\varphi^i\dot\varphi^h\xi^k \; , \quad (r = 1, 2, ..., n) \; ,$$

which is the desired result, i.e. an invariant form for the geodesic deviation.

**4**. – The system of eqs. (3.13) is *indeterminate* because the correspondence rule between the points *P* of *B* and *M* of γ is *arbitrary*. Indeed, it is easy to prove that the equation

(4.1) $$\frac{d}{d\sigma}\left[\lambda - \frac{D\xi^r}{D\sigma}\dot\varphi_r\right] = 0$$

is a consequence of eqs. (3.13). Accordingly, eq. (3.6) is a particular integral of eqs. (3.13); it specifies *one* of the integration constants. If we specialize our formalism to the case of pseudo-Riemannian geometry of general relativity, we see that the above arbitrariness yields a corresponding *vagueness* of the *physical* effects of a geodesic deviation. A common trick consists, first of all, in *assuming* that

(4.2) $$d\lambda/d\sigma = 0 \; , \quad \Rightarrow \quad \lambda = k \; , \quad \Rightarrow$$

(4.2′) $$\frac{ds}{d\sigma} = 1 + k \; ,$$

where *k* is an infinitesimal constant. Consequently

(4.3) $$\frac{D\xi^r}{D\sigma^2} = -R^r{}_{ihk}\,\dot\varphi^i\dot\varphi^h\xi^k \; ;$$

thus the arbitrariness has been formally reduced. Owing to eq. (4.1), the system (4.3) admits the integral

(4.4) $$\frac{D\xi^r}{D\sigma}\dot{\varphi}_r = k \ .$$

Now, by virtue of the following identity

(4.5) $$\frac{d}{d\sigma}\left(v^r\dot{\varphi}_r\right) = \frac{Dv^r}{D\sigma}\dot{\varphi}_r \ ,$$

where $v^r$ is any contravariant vector, we have

(4.6) $$\frac{D\xi^r}{D\sigma}\dot{\varphi}_r = \frac{d}{d\sigma}\left(\xi^r\dot{\varphi}_r\right) \ ;$$

therefore

(4.7) $$\xi^r\dot{\varphi}_r = k\sigma + C \ ,$$

where $C$ is another constant. If, in particular, $\lambda=k=0$, we obtain the relation

(4.8) $$\xi^r\dot{\varphi}_r = C \ ,$$

whose geometrical meaning is very simple. Finally, if we put $C=0$, the infinitesimal vector $PM$ becomes orthogonal to the base $B$, and any arbitrariness disappears: *par décret du physicien*.

**5**. – We shall now write eqs. (4.3) and equation

(5.1) $$\xi^r\dot{\varphi}_r = 0 \ ,$$

in a Riemann-Fermi system of geodesic co-ordinates $y^r$, ($r=1, 2, \ldots, n$). Correspondence rule: a given point $M$ of $\gamma$ corresponds to the point $P$ of $B$ with the same $y^n$ and the other $y^\alpha$, ($\alpha=1, 2, \ldots, n-1$), equal to zero. Thus the infinitesimal variations $\eta^r$ of the co-ordinates $y^r$ are:

(5.2) $$\eta^\alpha = y^\alpha, \ (\alpha=1, 2, \ldots, n-1) \ ; \ \eta^n = 0 \ .$$

Since $\sigma=y^n$, the parameters $\dot{\varphi}^\alpha = dy^\alpha/d\sigma$ of $B$ are equal to zero, while $\dot{\varphi}^n = dy^n/d\sigma = 1$; along the base $B$, the Christoffel symbols – and their first derivatives with respect to $\sigma$ – are equal to zero. Thus

(5.3) $$\frac{D\eta^r}{D\sigma} = \frac{d\eta^r}{d\sigma} \quad .$$

Accordingly, eqs. (4.3) can be written:

(5.4) $$\frac{d^2\eta^\alpha}{d\sigma^2} = -R^\alpha{}_{nn\beta}\eta^\beta \quad , \quad (\alpha, \beta=1, 2, \ldots, n-1).$$

(5.5) $$\frac{d^2\eta^n}{d\sigma^2} = -R^n{}_{nn\beta}\eta^\beta \quad ;$$

remark that $R^n{}_{nn\beta}$ is identically zero.

The astrophysicists intend to utilize eqs. (4.3), (5.1) – written for space-times of general relativity – for detecting the gravity waves. Typically, they hope to succeed in evidencing the effects of a geodesic deviation, generated by a wave, upon the suspended mirrors of their Michelson interferometers.

**6**. – The Newtonian analogue of eqs. (5.4), (5.5) is as follows. Consider two infinitely near trajectories of two test particles in a given gravitational field, whose potential is $U$. With reference to a Cartesian orthogonal system of axes $Ox_\alpha$, ($\alpha = 1, 2, 3$), the equations of motion are

(6.1) $$\frac{d^2 z_\alpha}{dt^2} = -\frac{\partial U}{\partial z_\alpha} \quad ; \quad \frac{d^2 Z_\alpha}{dt^2} = -\frac{\partial U}{\partial Z_\alpha} \quad ,$$

where $Z_\alpha = z_\alpha + \eta_\alpha$, and the $\eta_\alpha$ 's are the components of an infinitesimal vector. The relative acceleration is

(6.2) $$\frac{d^2\eta_\alpha}{dt^2} = -\frac{\partial U}{\partial z_\alpha \partial z_\alpha}\eta_\beta \quad .$$

**7**. − Finally, Levi-Civita [3] remarks that for *n*=2, if $y^2 = s$ is the curvilinear abscissa on *B*, and $y^1 = y$ is the distance between *M* and *B*, eqs. (5.4) , (5.5) are reduced to the unique equation

(7.1) $$\frac{dy^2}{d\sigma^2} = -R^1_{221}\ y \quad ;$$

remembering that in general, (*r*=1, 2, … ,n),

(7.2) $$R_{jihk} = a_{jr}R^r_{ihk} \quad ,$$

and that the Gaussian curvature *K* of a $V_2$ is given by

(7.3) $$\frac{R_{1212}}{\det\|a_{rs}\|} = \frac{R_{2121}}{\det\|a_{rs}\|} \quad ,$$

we see that eq. (7.1) coincides with Jacobi's formula (3.1):

(7.4) $$\frac{d^2 y}{d\sigma^2} + K(\sigma)y = 0 \quad ;$$

indeed, for our geodesic co-ordinates along *B* $\det\|a_{rs}\|=1$, and $R^1_{221} = R_{2121}$.

**8**. − All physical waves have an *independent* existence with respect to their source, which could even vanish after the emission of the wave. This should be true also for a hypothetical gravity wave. Now, the *undulatory nature* of such a "wave" is *frame dependent* [4]. However, even for a system of reference in which the wave character has disappeared, the curvature tensor remains obviously different from zero, provided it was different from zero in a previous frame.

Accordingly, there exists a *geodesic deviation*. I affirm that *no* effect can be produced by *this* geodesic deviation: indeed, the incoming of the gravity perturbation on the apparatus – resonant bar or Michelson interferometer – should provoke an increase of its *real* energy-momentum at the expense of the *pseudo* energy-momentum of the perturbation. But the pseudo energy-tensor "must be regarded as a mathematical fiction, not representing a significant condition of the world." [5]. It "can be created and destroyed at will by changes of co-ordinates; and even in a world containing no attracting matter (flat space-time) it does not necessarily vanish. It is therefore impossible to regard it as of a nature homogeneous with the proper energy-tensor." [5].

A similar argument holds for the celebrated case of the binary PSR1913+16, a system of two compact stars. It is commonly believed that the slow decrease of their revolution period is due to emission of gravitational waves. This is quite absurd because the *real* energy lost by the binary system in its revolutions should be transformed into the *pseudo* energy of the gravitational waves. The unjustifiable faith in the quadrupole formula has engendered this conceptual absurdity. (The above decrease can be brought about, *e.g.*, by viscosity phenomena).

Also the attempts to reveal the gravity waves by exploiting their interaction with an electromagnetic radiation are doomed to a failure: indeed, no fraction of the *real* energy-momentum of an e.m. wave can be transformed into the *pseudo* energy-momentum of a gravitational radiation.

*An obvious conclusion*: the gravity waves are mere phantoms.

**9**. – De Salvo writes (p.13 of his paper [1]): "LISA [the joint project of NASA in the US and ESA in Europe] is only sensitive to super-slow signals, orbital periods of many seconds or longer, from black holes falling into super-massive galactic black holes, and addresses a frequency band below the minimum achievable on the Earth's crust."

Now, I have proved with many arguments that the black holes are fictive objects, generated by an erroneous interpretation of the standard form of the solution to the problem of the Einstein field of a material point [6]. Here I shall limit myself to quote a significant sentence by Eddington, which too many physicists seem to ignore. After the deduction of the above standard form – *which is due to Hilbert, Droste, and Weyl* (quite independently), *and which is **different from the original** form of solution given by Schwarzschild* [7] – this Author examines the case of the isotropic co-ordinates [8]; at the end of sect.**43**, he writes: "The isotropic system could have been found directly by seeking particular solutions of Einstein's equations having the form […]

$$ds^2 = -\exp(\lambda)dr^2 - \exp(\mu)[r^2 d\theta^2 + r^2 \sin^2\theta \, d\varphi^2] + \exp(\nu)dt^2 \quad,$$

where $\lambda, \mu, \nu$ are functions of $r$. […]. Owing to an identical relation between $G_{11}$ [$\equiv R_{11}$], $G_{22}$ [$\equiv R_{22}$], and $G_{44}$ [$\equiv R_{44}$], the vanishing of this tensor gives only two equations to determine the three unknowns $\lambda, \mu, \nu$. There exists therefore an infinite series of particular solutions, differing according to the third equation between $\lambda, \mu, \nu$ which is at our disposal. The two solutions hitherto considered [i.e. the standard solution, commonly, but improperly, called "by Schwarzschild", and the isotropic one] are obtained by taking $\nu=0$ and $\lambda=\mu$, respectively. *The same series of solutions is obtained in a simpler way by substituting arbitrary functions of r instead of r in* (38.8) [*the standard form*]. [The italics are mine, A.L.].

The possibility of substituting any function of $r$ for $r$ without destroying the spherical symmetry is obvious from the fact that a co-ordinate is merely an identification–number; but analytically this possibility is bound up with the existence of an identical relation between $G_{11}$, $G_{22}$ and $G_{44}$, which makes the equations too few to determine a unique solution.".

Just by exploiting the above possibility, I have demonstrated the physical non-existence of the black holes [6]. (Frequently, the astrophysicists attribute to a black hole some effects that can be very well referred to a celestial body which collapsed in accordance with some refined *Newtonian* model [9]).

***Another obvious conclusion***: The statement that fictive objects (the black holes) can send forth fictitious entities (the gravity waves) contains a double conceptual error.